\documentclass[aps,prb,twocolumn,groupedaddress,showpacs]{revtex4-1}
\usepackage[english]{babel}
\usepackage{stmaryrd}
\usepackage{amssymb}
\usepackage{amsfonts}
\usepackage{amsmath}
\usepackage{color}
\usepackage{xspace}
\usepackage{graphicx}
\usepackage{bm}

\begin{document}

\title{Statistical properties of the one-dimensional Burridge-Knopoff model of earthquakes obeying the rate and state dependent friction law}

\author{Hikaru Kawamura}
\email{kawamura@ess.sci.osaka-u.ac.jp}

\author{Yushi Ueda}
\author{Shingo Kakui}
\author{Shouji Morimoto}
\author{Takumi Yamamoto}

\affiliation{Graduate School of Science, Osaka University,
Toyonaka, Osaka 560-0043, Japan}

\date{\today}

\begin{abstract}
Statistical properties of the one-dimensional spring-block (Burridge-Knopoff) model of earthquakes obeying the rate and state dependent friction law are studied by extensive computer simulations. The quantities computed include the magnitude distribution, the rupture-length distribution, the mainshock recurrence-time distribution, the seismic time correlations before and after the mainshock, the mean slip amount, and the mean stress drop at the mainshock, {\it etc\/}. Events of the model can be classified into two distinct categories. One tends to be unilateral with its epicenter located at the rim of the rupture zone of the preceding event, while the other tends to be bilateral with enhanced ``characteristic'' features resembling the so-called ``asperity''. For both types events, the distribution of the rupture length $L_r$ exhibits an exponential behavior at larger sizes, $\approx \exp[-L_r/L_0]$ with a characteristic ``seismic correlation length'' $L_0$. The mean slip as well as the mean stress drop tends to be rupture-length independent for larger events. The continuum limit of the model is examined, where the model is found to exhibit pronounced characteristic features. In the continuum limit, the characteristic rupture length $L_0$ is estimated to be $\sim $100 [km]. This means that, even in a hypothetical homogenous infinite fault, events cannot be indefinitely large in the exponential sense, the upper limit being of order $\sim 10^3$ kilometers. 
Implications to real seismicity are discussed. 
\end{abstract}

\maketitle

\section{Introduction}

 It has been realized for years that the scale-invariant power-law behaviors  are frequently observed in statistical properties of earthquakes, {\it i.e.\/}, the properties for an ensemble of earthquakes which are obtained after averaging over sufficiently many events. These include the well-known Gutenberg-Richter (GR) law  for the magnitude distribution of earthquakes and the Omori law for the time evolution of the frequency of aftershocks. Such an observation motivates the ``self-organized criticality (SOC)'' view of earthquakes \cite{Bak}, which regards the Earth's crust as being in the critical state which is self-generated dynamically \cite{Turcotte, Rundle2000, Hergarten, Rundle2003, Kanamori, Pradhan}. In contrast, one should also bear in mind that real earthquakes often exhibit apparently opposite features, {\it i.e.\/}, the features represented by  ``characteristic earthquakes'' where an earthquake is regarded to possess its characteristic energy or time scale \cite{Scholz2002, Kanamori, Ishibe}. 

 Statistical physical study of earthquakes is often based on simplified models of various levels of simplification \cite{Kawamura2012}. There are several advantages in employing simplified models in the study of earthquakes. In the model study, it is straightforward to control various material parameters as input parameters, whereas a systematic field study of the material-parameter dependence of real earthquakes meets serious difficulties. Furthermore, large earthquakes are rare, occurring, say, once hundreds of years for a given fault, and it is extremely difficult to examine the reproducibility of the observed phenomena and to take data with reliable error bars attached. In the model studies, on the other hand, it is often not difficult to put reliable error bars to the data under well controled conditions, say,  by performing extensive computer simulations. 

 One of the standard models widely employed in statistical physical study of earthquakes might be the Burridge-Knopoff (BK) model \cite{CLS, Kawamura2012}. The model was first introduced in Ref.\cite{BK}.  In the BK model, an earthquake fault is simulated by an assembly of blocks, each of which is connected via the elastic springs to to the neighboring blocks and to the moving rigid plate, and are slowly driven by an external force mimicking the plate drive.  As discussed in Ref.\cite{Ueda2014}, the many-block BK model may well represent the motion of a deformable fault layer, possibly corresponding to the low-velocity fault zone (LVFZ) observed in many long mature faults \cite{Huang}, pulled uniformly by the more or less rigid crust contingent to it. The model might also be useful in describing other stick-slip-type phenomena such as landslides \cite{ViescaRice}.

 In earthquakes studies, the simplicity of the BK model might provide us some benefitial points as compared with, {\it e.g.\/}, the standard elastodynamical continuum model. For one, the simplicity of the model often enables one to generate sufficiently many events, say, hundres of thousands of events, to reliably evaluate the statistical properties with reliable statistical precision, while, in the continuum model, only small number of events are usually generated. For the other, the BK model contains only small number of fundamental parameters, which makes it possible via the systematic survey of these parameter dependence to concentrate on the role and the inter-relation of these small number of fundamental parameters in earthquake occurrence, to extract and clarify the physical mechanism underlying apparently complex earthquake phenomena. 
Finally, the BK model, even in its continuum limit, describes a setting a bit different from that of the standard elastodynamical continuum model.

 A crucially important part of the model might be the type of friction force assumed \cite{Scholz1998, Scholz2002}.  In the pioneering study of the statistical properties of the BK model, Carlson, Langer and collaborators employed the simple velocity-weakening friction force. The friction force is assumed to be a single-valued decreasing function of the velocity \cite{CL1989a, CL1989b, CLST,Carlson1991a, Carlson1991b, Shaw, CLS, Schmittbuhl, MoriKawamura2005, MoriKawamura2006, MoriKawamura2008a, MoriKawamura2008b, MoriKawamura2008c, Kawamura2012}.

 More realistic constitutive relation now standard in seismology might be the rate-and-state dependent friction (RSF) law \cite{Dieterich, Ruina, Marone}. The RSF law assumes that the friction depends not only on the slip velocity $V$ but also on the ``state'' of the slip interface, which is phenomenologically described via the ``state variable'' $\Theta$ obeying its own evolution law. The time-evolution law of the state variable generally includes a characteristic slip distance ${\mathcal L}$, which gives a measure of the length scale at which a slip interface loses its initial memory of the state.
 
 This RSF law has widely been used in numerical simulations mostly on the continuum model \cite{TseRice, Stuart, Horowitz, Rice, BenZionRice, KatoHirasawa, Kato, Bizzarri2006a, Bizzarri2006b}, but also on the BK model. For example,  Cao and Aki performed a numerical simulation of earthquakes by combining the 1D BK model with the RSF law in which various constitutive parameters were set nonuniform over blocks \cite{CaoAki}. Ohmura and Kawamura extended an earlier calculation by Cao and Aki to study the statistical properties of the 1D BK model combined with the RSF law with uniform constitutive parameters \cite{OhmuraKawamura, Kawamura2012}. Clancy and Corcoran also performed a numerical simulation of the 1D BK model based on a modified version of the RSF law \cite{Clancy}. 

 Of course, the space discretization in the form of blocks inherent to the BK model is an approximation to the original continuum crust. (Note, however, that the discreteness may also be regarded as a measure of the underlying spatial inhomogeneity \cite{Rice}.) It introduces the short-length cut-off scale into the problem in the form of the block size, which could in principle give rise to an artificial effect not realized in the continuum.

 Rice criticized that the discrete BK model with a simple velocity-weakening law was ``intrinsically discrete'', lacking  in a well-defined continuum limit, arguing that the spatiotemporal complexity observed in the discrete BK model was due to an inherent discreteness of the model, which should disappear in continuum \cite{Rice, BenZionRice}. If the grid spacing $d$ is taken larger than  the ``nucleation length'' which was proportional to the slip distance ${\mathcal L}$, the system exhibits an apparently complex or critical behavior, whereas, if the grid spacing $d$ is taken smaller than it, the system tends to exhibit a quasi-periodic recurrence of large events. In this picture, the  block-discretization effect of the BK model should closely be related to its nucleation phenomena via the nucleation length. In the continuum limit where the grid spacing tends to zero, the system is expected to always exhibit a quasi-periodic or a ``characteristic'' behavior where large earthquakes repeat near-periodically without critical features \cite{Rice}. 

 Indeed, we recently examined the nature of the nucleation process of the discrete BK model under the RSF law by systematically varying the extent of the discreteness of the model toward the continuum limit, and also by systematically varying various model parameters including the frictional and the elastic parameters \cite{Ueda2014,Ueda2015}. It was observed that the model exhibited a quasi-static initial phase in its nucleation process when the frictional instability was weak, {\it i.e.\/}, when the normalized frictional-weakening parameter $b$ was less than a critical value $b_c$ determined by the elastic-stiffness parameter $l$ as $b_c=2l^2+1$, while the quasi-static initial phase was absent when the frictional instability was strong, {\it i.e.\/}, when $b>b_c=2l^2+1$. The continuum limit entails the relation $l\rightarrow \infty$ so that the continuum limit of the BK model under the RSF law necessarily lies in the weak frictional instability regime, and accompanies a long-durating quasi-static nucleation process.

 In view of these recent findings on the nucleation phenomena of the 1D BK model under the RSF law, we wish to examine in the present paper the statistical properties of subsequent mainshocks themselves, by systematically varying the extent of the model discreteness and various model parameters, paying particular attention to its {\it characteristic\/} versus {\it critical\/} features. The computation of the present paper is an extension of the earlier calculation of Ref.\cite{OhmuraKawamura} on the same model. These authors concentrated on the strong frictional instability regime, studying the limited number of observables, {\it i.e.\/}, the magnitude distribution and the recurrence-time distribution. In the present paper, we deal with not only the strong frictional instability regime but also the weak frictional instability regime, even including its continuum limit. Note that the weak frictional instability regime is computationally more demanding since it necessarily accompanies a slow nucleation process which requires more computational resources. We also compute various observables not computed in Ref.\cite{OhmuraKawamura}, including the rupture-zone size (rupture length) distribution, the seismic time correlations before and after the mainshock, the mean slip amount and the mean stress drop at the mainshock, {\it etc\/}, aimed at reaching deeper understanding of the the nature of seismic events of the model. 
 
  We then find that the characteristic feature of mainshocks becomes more pronounced as one moves from the strong to the weak frictional instability regime. While the magnitude distribution of the model in the strong frictional instability regime of $b>b_c$ exhibits an almost flat distribution spanning from smaller to larger events as reported in  \cite{OhmuraKawamura}, the distribution tends to be more peaked at a characteristic magnitude as one moves to the weak frictional instability regime of $b<b_c$. It means that a hypothetical, uniform fault obeying the RSF law tends to exhibit a pronounced characteristic behavior, accompanied by the quasi-periodic recurrence of earthquakes of more or less similar magnitude. Such a characteristic property is in apparent contrast to the power-law critical behavior as embodied by the GR law, but corroborates the Rice's claim \cite{Rice}.

 When one looks at the rupture-length ($L_r$) distribution, a simpler behavior turns out to emerge. Both in the strong and the weak instability regimes, the distribution exhibits an exponential behavior at larger sizes, $\approx \exp[-L_r/L_0]$, characterized by the characteristic ``{\it seismic correlation length\/}'' $L_0$. This observation hints that certain forecast might be possible, at least for a mature homogenous fault, on the basis of such pronounced characteristic features. It is also observed that not only the mean stress drop but also the mean slip amount at the mainshock tends to be $L_r$-independent for larger events.

 Events of the model might be classified into two distinct categories, called here the type-I and the type-II events. The type-I event tends to occur with its epicenter located at the rim of the rupture zone of the preceding event, and tends to be unilateral, {\it i.e.\/}, its rupture propagates predominantly into one direction and the epicenter lies near the edge of the rupture zone of the event. By contrast, the type-II event tends to occur with its epicenter located in the interior of this event, the rupture propagating into both directions. The type-II event has an enhanced characteristic feature than the type-I event. For example, the type-II event tends to repeat several times with a more or less common epicenter and rupture zone, with features of the so-called ``asperity''. The dominance of either type-I/II events depends on the weak/strong frictional instability regime, {\it i.e.\/}, the type-I (type-II) event tends to dominate in the strong (weak)  frictional instability regime. The fact that seismic events of the model tend to be increasingly more characteristic in the weaker frictional instability regime might be understood from the dominance of the type-II event in the weak frictional instability regime.

 Continuum limit of the model is also examined. Since the continuum limit of the model always lies in the weak frictional instability regime of the original discrete model irrespective of its parameter values, our results suggest that seismic events of a mature homogeneous fault should be more or less ``characteristic'', with features of asperities. Such enhanced characteristic features enable one to discuss about ``typical scales'' underlying the seismic events, those of length, time and energy. We then try to give explicit estimates of these scales underlying seismicity.

 Overall, the properties of the BK model under the RSF law are sometimes considerably different from those of the BK model under the pure velocity-weakening law employed in most of the previous simulations on the BK model \cite{CL1989a, CL1989b, CLST,Carlson1991a, Carlson1991b, Shaw, CLS, Schmittbuhl, MoriKawamura2005, MoriKawamura2006, MoriKawamura2008a, MoriKawamura2008b, MoriKawamura2008c, Kawamura2012}. Roughly speaking, characteristic features tend to be more enhanced in the RSF-law model than in the pure velocity-weakening-law model. 

 The rest of the paper is organized as follows. In section II, we define our model, the 1D BK model obeying the RSF law, and present its equation of motion. In section III, we report the results of our numerical simulation on various statistical properties of the model, {\it e.g.\/}, the magnitude distribution, the rupture-length distribution, the mainshock recurrence-time distribution, the seismic time correlations before and after the mainshock, the mean slip amount and the mean stress drop at the mainshock, {\it etc\/}, covering both the strong and the weak instability regimes. In section IV, we deal with the continuum limit of the model, and investigate how various statistical properties behave in this limit. Finally, section V is devoted to summary and discussion. Implications to real seismicity are discussed.

\section{The model}

 The 1D BK model consists of a 1D array of $N$ identical blocks of the mass $m$, which are mutually connected with the two neighboring blocks via the elastic springs of the spring stiffness $k_c$, and are also connected to the moving plate via the springs of the spring stiffness $k_p$, and are driven with a constant rate $\nu'$. All blocks are subject to the friction strength $\Phi$, which is the source of nonlinearity in the model. The equation of motion for the $i$-th block can be written as
\begin{equation}
m \frac{{\rm d}^2U_i}{{\rm d}t^{\prime 2}} = k_p (\nu ' t'-U_i) + k_c (U_{i+1}-2U_i+U_{i-1})-\Phi_i,
\label{original-eq-motion}
\end{equation}
where $t'$ is the time, $U_i$ is the displacement of the $i$-th block, and $\Phi_i$ is the friction force at the $i$-th block. For simplicity, the  motion in the direction opposite to the plate drive is inhibited by imposing an infinitely large friction for $\dot U_i<0$.

For the friction law, we assume the RSF law given  by 
\begin{eqnarray}
\Phi_i=\left\{C + A \log(1+\frac{V_i}{V^*}) + B \log \frac{V^* \Theta_i}{{\mathcal L}} \right\} {\mathcal N} ,
\label{original-RSF}
\end{eqnarray}
where $V_i=\frac{{\rm d}U_i}{{\rm d}t^\prime}$ is the velocity of the $i$-th block, $\Theta_i(t')$ is the time-dependent state variable (with the dimension of time) representing the ``state'' of the slip interface, $V^*$ is a crossover velocity underlying the RSF law, ${\mathcal N}$ is an effective normal load, ${\mathcal L}$ is a critical slip distance which is a measure of the sliding distance necessary for the surface to evolve to a new state, with $A,\ B$ and $C$ positive constants describing the RSF law. The first term ($C$-term) is a constant taking a value around $\frac{2}{3}$ \cite{Scholz2002}, which dominates the total friction in magnitude, the second term ($A$-term) a velocity-strengthening direct term describing the part of the friction responding immediately to the velocity change, the third part ($B$-term) an indirect velocity-weakening term dependent of the state variable. Laboratory experiments suggest that the $A$- and $B$-terms are smaller than the $C$-term by one or two orders of magnitudes, yet they play an essential role in stick-slip dynamics \cite{Marone, Scholz1998, Scholz2002}. 

 Note that, in the standard RSF law, the $A$-term is often assumed to be proportional to $\log(\frac{V_i}{V^*})$. Obviously, this form becomes pathological in the $V\rightarrow 0$ limit because it gives a negatively divergent friction. In other words, the pure logarithmic form of the $A$-term cannot describe the state at a complete rest. We cure this pathology by phenomenologically introducing a modified form given above \cite{Ueda2015}. The modified form, where the $A$-term becomes proportional to the block velocity $V$ at $V<<V^*$ but reduces to the purely logarithmic form at $V>>V^*$, is enable to describe a complete halt. The characteristic velocity $V^*$ represents a crossover velocity, describing a low-velocity cutoff of the logarithmic behavior of the friction.
 
 For the evolution law of the state variable, we use here the so-called aging (slowness) law given by
\begin{eqnarray}
\frac{d\Theta_i}{dt'}=1-\frac{V_i\Theta_i}{{\mathcal L}}.
\label{original-aging}
\end{eqnarray}
Under this evolution law, the state variable $\Theta_i$ grows linearly with time at a complete halt $V_i=0$ reaching a very large value at the outset of the nucleation process, while it decays very rapidly during the seismic rupture.

 The equation of motion can be made dimensionless by taking the length unit to be the critical slip distance ${\mathcal L}$, the time unit to be $\omega^{-1}=\sqrt{m/k_p}$ and the velocity unit to be $\mathcal{L}\omega$,
\begin{eqnarray}
\frac{{\rm d}^2u_i}{{\rm d}t^2} &=& \nu t-u_i+l^2(u_{i+1}-2u_i+u_{i-1}) \nonumber \\ 
&-& \left( c+a\log \left(1+\frac{v_i}{v^*}\right)+b\log \theta_i \right) ,
\label{eq-motion} \\
\frac{{\rm d}\theta_i}{{\rm d}t} &=& 1-v_i\theta_i ,
\label{aging}
\end{eqnarray}
where the dimensionless variables are defined by $t=\omega t'$, $u_i=U_i/{\mathcal L}$,  $v_i=V_i/(\mathcal{L}\omega)$, $v^*=V^*/(\mathcal{L}\omega)$, $\nu=\nu'/(\mathcal{L}\omega)$,  $\theta_i=\Theta_i \omega$, $a=A{\mathcal N}/(k_p{\mathcal L})$, $b=B{\mathcal N}/(k_p{\mathcal L})$, $c=C{\mathcal N}/(k_p{\mathcal L})$, while $l \equiv \sqrt{k_c/k_p}$ is the dimensionless elastic stiffness 
parameter. 

 It is sometimes more convenient to rewrite the equation of motion in terms of the velocity variable $v_i$ instead of the displacement $u_i$. By differentiating (\ref{eq-motion}) with respect to $t$ and by using (\ref{aging}), one gets
\begin{eqnarray}
\frac{d^2v_i}{dt^2} &+&  \frac{a}{v_i+v^*}\frac{dv_i}{dt} + \left\{l^2(v_{i+1}-2v_i+v_{i-1})+1-b \right\} v_i \nonumber \\ 
 &=& \nu - \frac{b}{\theta_i} .
\label{eq-motion2}
\end{eqnarray}
The block displacement or the slip amount $u_i$ can be obtained up to a constant by integrating the velocity $v_i$ with respect to $t$.

 One sees from eqs.(\ref{eq-motion2}) and (\ref{aging}) that the constant frictional parameter $c$ no longer remains in the governing equations, meaning this parameter is essentially irrelevant to the dynamical properties of the model. In our simulations, we use either eqs.(\ref{eq-motion}) or (\ref{eq-motion2}) depending on the situation. In solving the high-speed motion, we use eq.(\ref{eq-motion}), while in solving the low-speed motion as realized in the initial phase or the early stage of the acceleration phase, we use eq.(\ref{eq-motion2}).

 The frictional parameter $a$/$b$ tends to suppress/enhance the frictional instability. The earthquake instability is driven primarily by the velocity-weakening $b$-term, while the velocity strengthening $a$-term tends to mitigate the unstable slip toward the aseismic slip. Since the frictional parameters $a$ and $b$ compete in their function, either $a<b$ or $a>b$ might affect the dynamics significantly. When $a\gtrsim b$, the compensation effect due to the $a$-term tends to induce a slow slip succeeding a mainshock, {\it i.e.\/}, an afterslip, while, when $a>>b$, it gives rise to slow-slip events (SSE), no longer accompanying the high-speed rupture at any stage of the event. Earthquake properties in this regime of $a>b$ will be reported in a separate paper, with emphasis on the slow-slip phenomena.

 Estimates of typical values of the model parameters representing natural earthquake faults have been given in Ref.\cite{Ueda2014}. The BK model and its continuum limit possess a built-in time scale, $\omega^{-1}$, corresponding to the typical rise time of an earthquake event, which may be estimated to be $\sim 1$ [sec]. The model possesses two distinct and independent length scales: one associated with the fault slip and the other with the distance along the fault. The former length scale is the critical slip distance ${\mathcal L}$, which was estimated to be $\simeq 1$ [cm], while the other is the distance the rupture propagates per unit time, $v_s/\omega$, which was estimated to be $\simeq 2-3$ [km] where $v_s$ is the $s$-wave velocity in the LVFZ. The spring constant $k_p$ was related to the normal stress as $\frac{{\mathcal N}}{k_p{\mathcal L}}\simeq 10^2-10^3$. Then, as $C$ is known to take a value around $\frac{2}{3}$, $c$ would be of order $10^2$-$10^3$, $a$ and $b$ being one or two orders of magnitude smaller than $c$. The crossover velocity $V^*$ and its dimensionless counterpart $v^*$ is hard to estimate though it should be much smaller than unity, and we take it as a parameter.

 The continuum limit of the BK model corresponds to making the dimensionless block size $d$ to be infinitesimal $d\rightarrow 0$, simultaneously making the system infinitely rigid $l=\frac{1}{d} \rightarrow \infty$ \cite{MoriKawamura2008c}. The dimensionless distance $x$ between the block $i$ and $i^{\prime}$ is $x=|i-i^{\prime}|d=\frac{|i-i^{\prime}|}{l}$.

 Since the continuum limit entails $l\rightarrow \infty$, the condition of the weak frictional instability $b<b_c=2l^2+1$ is always satisfied there. Hence, the continuum limit of the model always lies in the weak frictional instability regime accompanying the quasi-static nucleation process.

 Simulations are made by numerically solving the coupled equtions of motions for $v_i$ (or $u_i$) and $\theta_i$ ($1\leq i\leq N$) by means of the fourth-order Runge-Kutta method. The total number of blocks $N$ is taken to be $N=800$ in most cases, while other sizes up to $N=1600$ are studied to check the possible finite-size effects. Open boundary conditions are adopted for the block at each end of the system. We have checked that the results shown below for the systems size $N=800$ are free from finite-size effects in that the results have well converged against further increase of $N$. Concerning the initial conditions, all blocks are assumed to be ar rest, {\it i.e.\/}, $v_i=0$ ($1\leq i\leq N$) at $t=0$, the state variable is taken to be uniform $\theta_i=10^8$, while the displacement of each block is assumed to take random values uniformly distributed between -5 and 5 from block to block. Events at earlier times are just transient and non-stationary, strongly affected by the initial conditions. We wait until the system reaches the stationary state loosing initial memory, and compute various observables in such stationary states.

 We emphasize that, although the model is completely uniform or `homogenous' in the model parameters describing its equations of motion, it exhibits quite erratic or irregular behavior as will be shown in subsequent sections.  In other words, the outcoming state of the model, {\it e.g.\/}, the displacement, the velocity and the stress {\it etc.\/}, could be quite `inhomogeneous'. The origin of such irregular behaviors lies in the imposed initial conditions inevitably existing in any real setting, from which the irregularity or the complexity is self-evolved via the intrinsic ``chaotic'' dynamics.

\section{The results}

 In this section, we study various statistical properties of the 1D BK model, including the magnitude distribution, the rupture-length distribution, the recurrence time distribution, the mean slip amount, and the mean stress drop at the mainshock.

 We begin with the magnitude distribution. The magnitude of an event, $\mu$, may be defined by
\begin{equation}
\mu = \ln \left( \sum_i \Delta u_i \right),
\end{equation}
where the sum is taken over all blocks involved in the event.

\begin{figure}[ht]
\begin{center}
\includegraphics[scale=0.85]{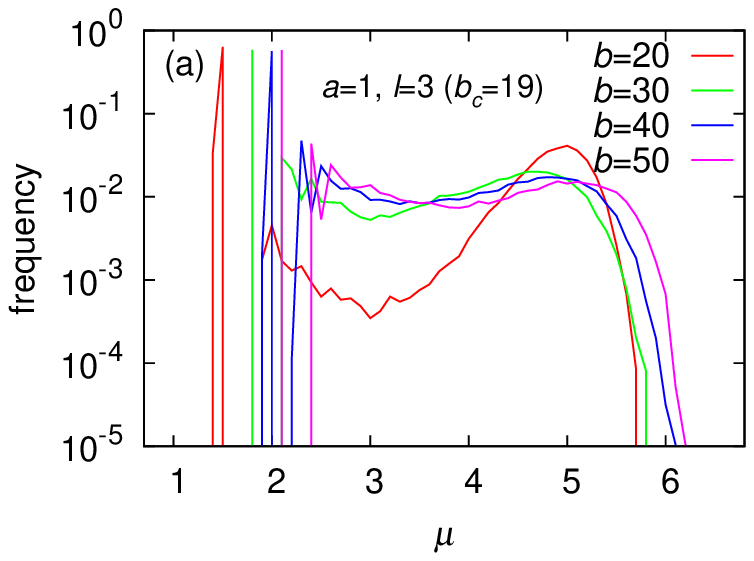}
\includegraphics[scale=0.85]{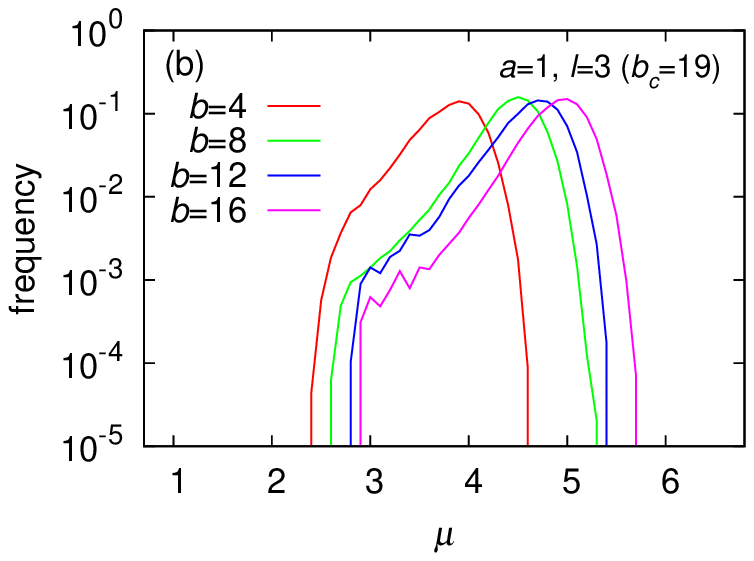}
\end{center}
\caption{
The magnitude distribution of earthquake events of the 1D BK model for various $b$-values; (a) in the strong frictional instability regime of $b>b_c=19$, and (b) in the weak frictional instability regime of $b<b_c$. The other parameters values are $a=1$, $l=3$, $c=1000$, $v^*=10^{-2}$ and $\nu=10^{-8}$. The system size is $N=800$.
}
\end{figure}

 The computed magnitude distribution is shown in Fig.1 both in the strong frictional instability regime of larger $b$ (a), and in the weak frictional instability regime of smaller $b$ (b). The parameters are taken to be $a=1$, $c=1000$, $l=3$, $\nu=10^{-8}$ and $v^*=10^{-2}$ so that the critical value of $b$ discriminating the weak/strong frictional instability is $b_c=2l^2+1=19$. As can be seen from Fig.1(a), the magnitude distribution in the strong frictional instability regime exhibits an almost flat distribution spanning from smaller to larger events. While events of various sizes tend to occur, the distribution does not obey the GR law. The result in the strong frictional instability regime is consistent with the earlier result of Ref.\cite{OhmuraKawamura}.

 As $b$ is decreased toward $b_c$ and the system approaches the weak frictonal instability regime, the magnitude distribution changes its shape with a more weight at a larger magnitude. In the weak frictional instability regime of $b<b_c$, the data exhibit a more characteristic behavior with a pronounced peak at a magnitude $\mu=\mu_p$, meaning large events of the magnitude $\mu_p$ predominantly occur. Hence, a mainshock in the weak frictional instability regime, which covers the continuum limit of the model, tends to acquire a pronounced characteristic feature. 

\begin{figure}[ht]
\begin{center}
\includegraphics[scale=0.85]{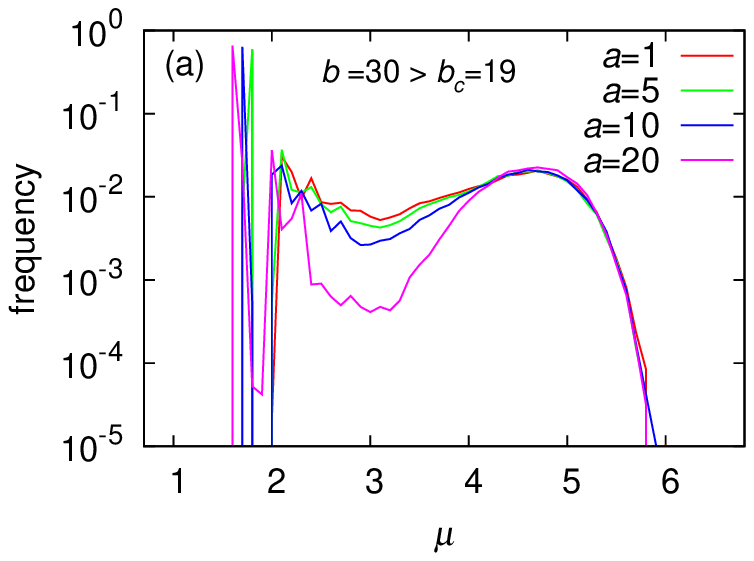}
\includegraphics[scale=0.85]{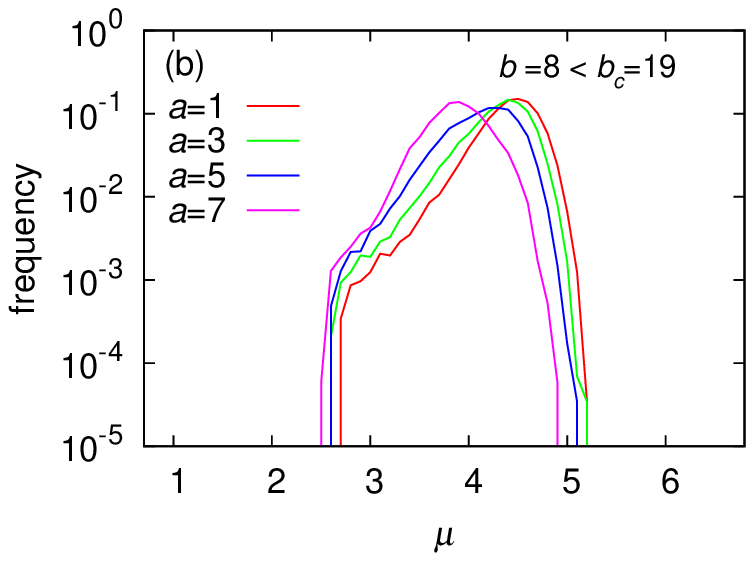}
\end{center}
\caption{
The magnitude distribution of earthquake events of the 1D BK model for various $a$-values; (a) in the strong frictional instability regime of $b=30>b_c=19$, and (b) in the weak frictional instability regime of $b=8<b_c$. The other parameter values are $l=3$, $c=1000$, $v^*=10^{-2}$ and $\nu=10^{-8}$. The system size is $N=800$.
}
\end{figure}

 We also examine the other parameter dependence of the magnitude distribution, the $a$-dependence in Fig.2, and the $v^*$-dependence in Fig.3 both in the strong and the weak instability regimes. As can be seen from these figures, the distribution hardly depends on both $a$ and $v^*$.  As can be seen from eq.(6), the magnitude distribution is also insensitive to the frictional parameter $c$, which we also confirmed (the data not shown here). 

\begin{figure}[ht]
\begin{center}
\includegraphics[scale=0.85]{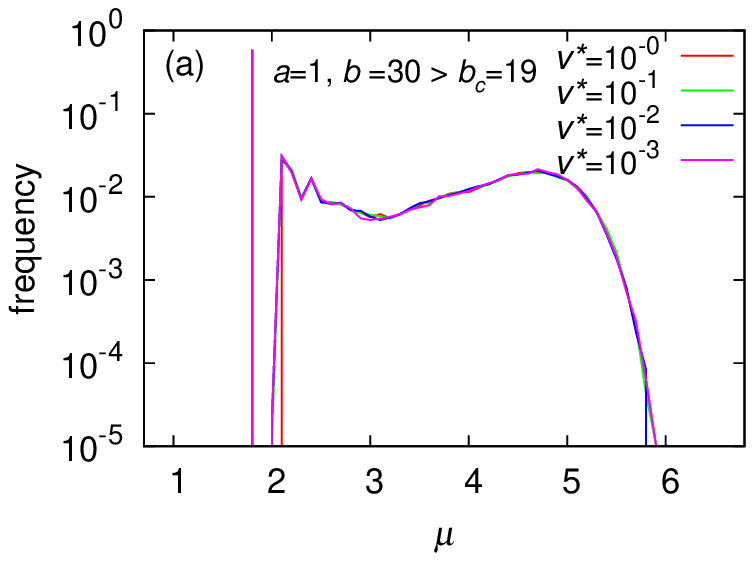}
\includegraphics[scale=0.85]{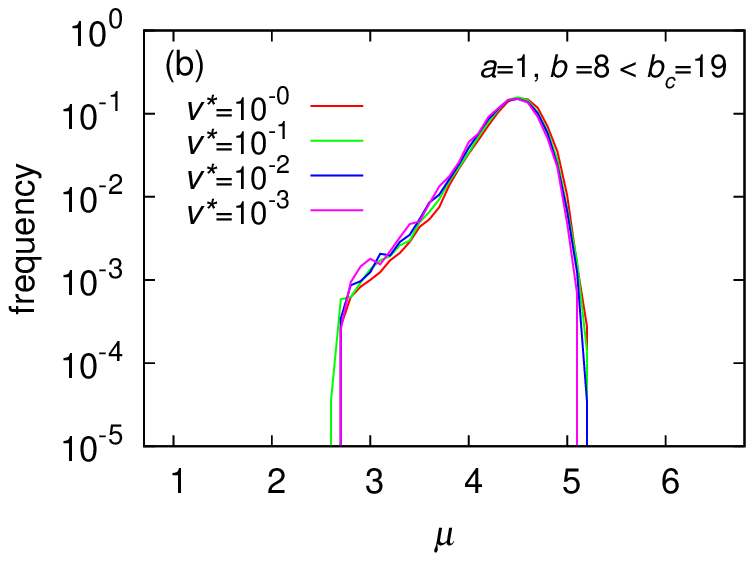}
\end{center}
\caption{
The magnitude distribution of earthquake events of the 1D BK model for various $v^*$-values; (a) in the strong frictional instability regime of $b=30>b_c=19$, and (b) in the weak frictional instability regime of $b=8<b_c$.  The other parameter values are $a=1$, $l=3$, $c=1000$ and $\nu=10^{-8}$. The system size is $N=800$.
}
\end{figure}

 A magnitude $\mu$ has been defined as the multiple of the rupture length $L_r$ and the mean slip amount $\bar u$.  If one looks at the rupture-length $L_r$ distribution, an interesting tendency shows up. In Fig.4, we show the distribution of $L_r$ on a semi-logarithmic plot for various $b$ in the strong (a) and in the weak (b) frictional instability cases. The parameter choice is the same as in Fig.1. As can be seen from Fig.4, the data tend to lie on a straight line except for smaller events, indicating that the distribution has an exponential form $\simeq \exp[-(L_r/L_0)]$ with a characteristic rupture length $L_0$. Such an exponential behavior prevails both in the weak and in the strong frictional instability regimes. The observed finite $L_0$ is not a finite-size effect. In the region where the mean slip amount $\bar u$ is nearly constant, which is the case for larger events in the weak frictional instability regime as will be shown in Fig.10 below, apparent different shapes between the $\mu$-distribution (Fig.1(b)) and the $L_r$-distribution (Fig.4(b)) might be understandable by noting the relation d$\mu \simeq {\rm d}\ln(\bar u L_r) \simeq (\bar u/L_r) {\rm d}L_r$.

\begin{figure}[ht]
\begin{center}
\includegraphics[scale=0.85]{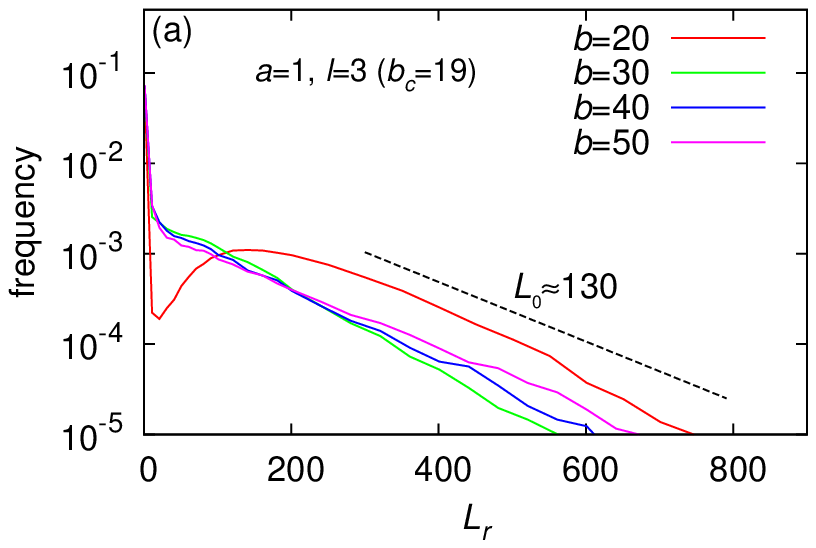}
\includegraphics[scale=0.85]{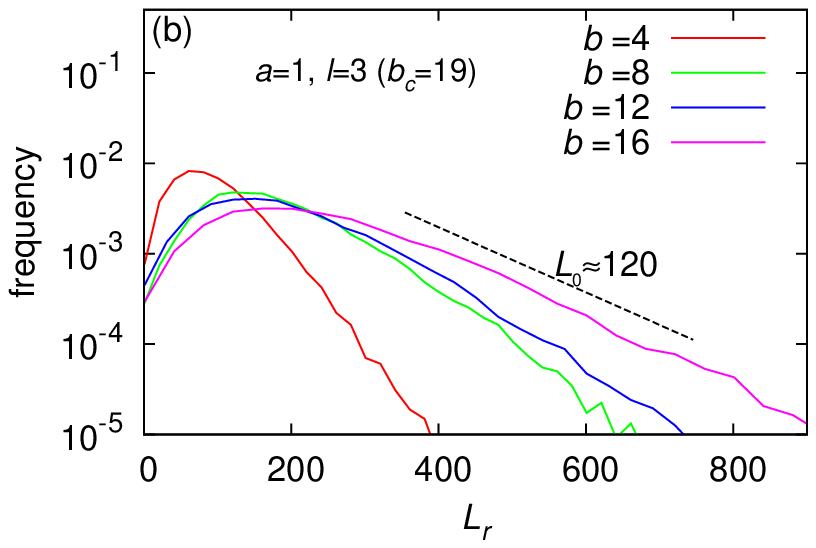}
\includegraphics[scale=0.85]{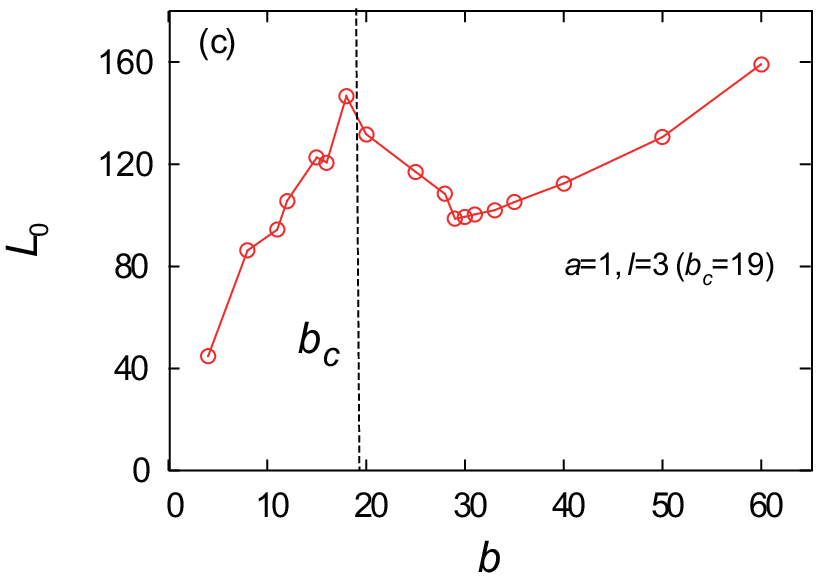}
\end{center}
\caption{
The rupture-length $L_r$ distribution of earthquake events  of the 1D BK model on a semi-logarithmic plot for various $b$-values; (a) in the strong frictional instability regime of $b>b_c=19$, and (b) in the weak frictional instability regime of $b<b_c$. The other parameter values are $a=1$, $l=3$, $c=1000$, $v^*=10^{-2}$ and $\nu=10^{-8}$. The system size is $N=800$. The inverse slope of the tail of the distribution yields a characteristic rupture-length scale $L_0$ as indicated in the figure. The $b$-dependence of $L_0$ is shown in Fig.(c).
}
\end{figure}

 We also examine the $b$-dependence of $L_0$, and the result is shown in Fig.4(c). At $b=b_c=19$ discriminating the weak/strong instability regimes, there occurs a change of behavior in the $b$-dependence of $L_0$. Anyway, the existence of a characteristic rupture length of the length scale $L_0\simeq 10^2$ seems to be a notable feature of the 1D BK model under the RSF law. We shall discuss its continuum limit in the next section. A change of behavior of the $b$-dependence of $L_0$ can be seen in Fig.4 also around $b\simeq 30$. This change of behavior is closely related to the observation that the dominant type of events changes around $b\simeq 30$. We shall return to this issue in Fig.9 below.

\begin{figure}[ht]a
\begin{center}
\includegraphics[scale=0.85]{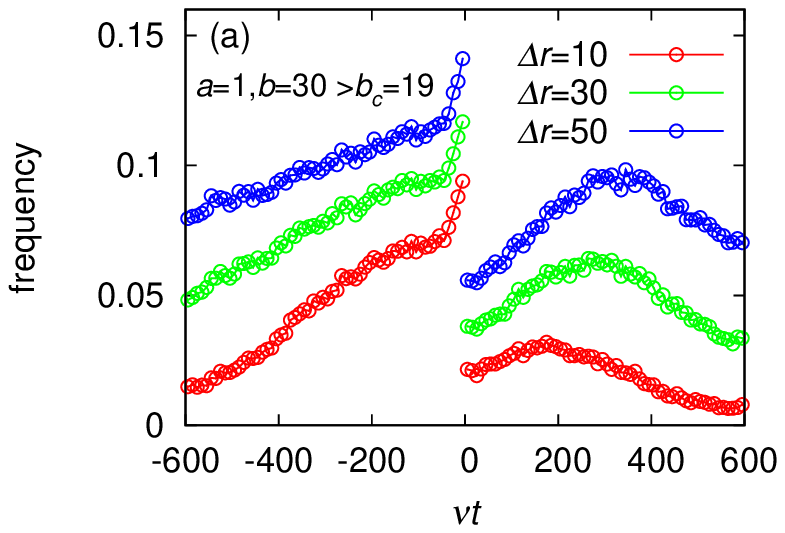}
\includegraphics[scale=0.85]{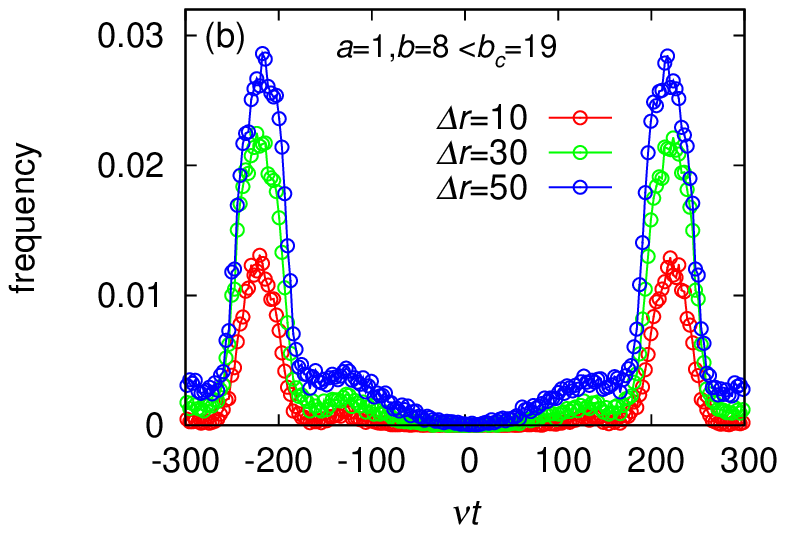}
\end{center}
\caption{
The event frequency before and after the mainshock occurring at the time $t=0$; (a) in the strong frictional instability regime of $b=30>b_c=19$, and (b) in the weak frictional instability regime of $b=8<b_c$. The other parameter values are $a=1$, $l=3$, $c=1000$, $v^*=10^{-2}$ and $\nu=10^{-8}$. The system size is $N=800$. The number of events occurring within the distance $\Delta r$ from the epicenter site of the mainshock is counted irrespective of their magnitude.
}
\end{figure}

 In real seismicity, a mainshock often accompanies aftershocks and foreshocks, which obey the Omori law or the inverse Omori law. For the BK model under the velocity-weakening friction law, by contrast, earlier studies indicated that such an aftershock (foreshock) sequence obeying the Omori (inverse Omori) law has hardly been observed \cite{MoriKawamura2006, MoriKawamura2008a, MoriKawamura2008b, MoriKawamura2008c}. Then, we investigate here the corresponding properties for the BK model under the RSF law. In Fig.5, the frequency of events correlated with mainshocks is shown as a function of the time $t$ both before ($t<0$) and  after ($t>0$) the mainshock. To make comparison with the previous works, we take here the definition of aftershocks/foreshocks same as those of Refs.\cite{MoriKawamura2006, MoriKawamura2008a, MoriKawamura2008b, MoriKawamura2008c}, {\it i.e.\/}, mainshocks are taken as events of their magnitude greater than $\mu_c=4$, and the frequency of all events which occur at time $t$ in the neighborhood of the epicenter of a mainshock, its epicenter being located within the distance $\Delta r$ (in units of block number) from the mainshock epicenter block, is plotted versus the time $t$. Average is made over mainshocks where the time origin $t=0$ is set common.

\begin{figure}[ht]
\begin{center}
\includegraphics[scale=0.85]{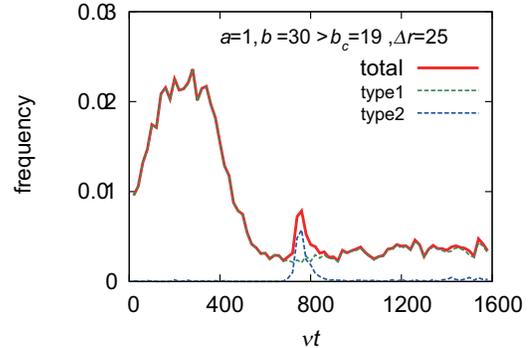}
\end{center}
\caption{
The distribution of the mainshock recurrence time of the 1D BK model. The parameter values are  $a=1$, $b=30$, $l=3$, $c=1000$, $v^*=10^{-2}$ and $\nu=10^{-8}$. The system size is $N=800$. The dashed lines represent the recurrence-time distribution of the type-I events (green) and of the type-II events (blue). For the details, see the main text.
}
\end{figure}

 The computed time correlation is shown in Fig.5 both in the strong (a) and in the weak (b) instability cases. The peak structure observed around $t\simeq 200\sim 300$ in Fig.5(a) and $t\simeq \pm 250$ in Fig.5(b) are associated with the subsequent (or preceding) mainshock. As can be seen from Figs.5(a) and (b), seismic activity tends to be suppressed after the mainshock, an aftershock sequence being not evident.

 In the strong frictional instability case of Fig.5(a), seismic activity tends to be gradually enhanced before the mainshock toward the mainshock. This enhancement occurs on the time scale of the recurrence period of mainshocks, representing a long-term activation of seismicity toward the mainshock rather than standard foreshocks. In the weak frictional instability regime, even such a long-term activation is not discernible. The computed time correlation takes a symmetric form before and after the mainshock, without standard foreshocks and aftershocks. As such, the standard aftershock/foreshock sequence obeying the Omori (inverse Omori) law is not realized in the 1D BK model even under the RSF law. Some ingredients not taken into account in the present model, {\it e.g.\/}, the higher-dimensionality effect and/or the slow relaxation process, seem to be necessary to realize foreshock/aftershock sequences.

 The interval (or the recurrence) time $T$ between mainshocks is also of interest. Ohmura and Kawamura studied the recurrence-time distribution of the model in the strong frictional instability case, and observed that it possessed a double-peak structure, each peaked at apparently independent times $T=T_1$ and $T_2$ ($T_1<T_2$). In order to get further insight into the issue, we compute here the local recurrence-time distribution in the strong instability case, and the result is shown in Fig.6. Mainshocks are defined here with their magnitude of $\mu\geq \mu_c=4$, while the mainshocks with its epicenter lying in the neighborhood of the preceding mainshock with $\Delta r\leq 25$ are counted as the next event. As can be seen from Fig.6, the double-peak structure is discernible at $T_1\simeq 300$ and $T_2\simeq 750$, though in a less pronounced compared with the ones observed in Ref.\cite{OhmuraKawamura}, presumably due the smaller value of $v^*$ adopted here, {\it i.e.\/}, $v^*=10^{-2}$ here versus $v^*=1$  in \cite{OhmuraKawamura}.

\begin{figure}[ht]
\begin{center}
\includegraphics[scale=0.85]{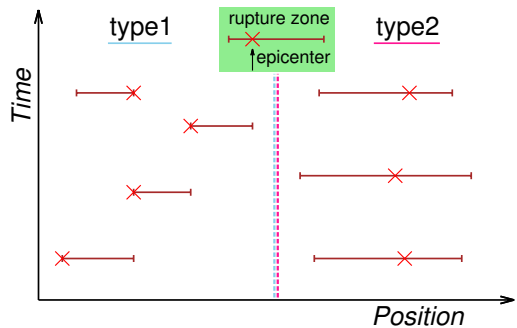}
\end{center}
\caption{
Schematic spatiotemporal pattern of the type-I and the type-II events. Bars represent the rupture zone of the event, and crosses represent its epicenter site.
}
\end{figure}
\begin{figure}[ht]
\begin{center}
\includegraphics[scale=0.85]{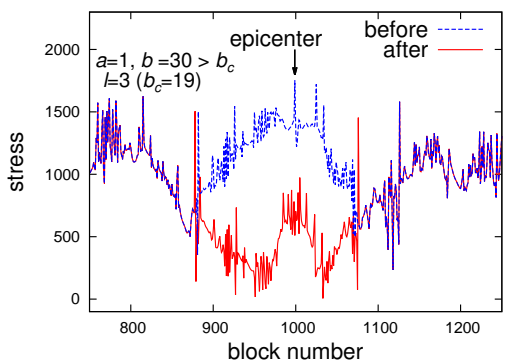}
\end{center}
\caption{
The stress value plotted versus the block position just before and after a typical type-II event. The parameter values are  $a=1$, $b=30$, $l=3$, $c=1000$, $v^*=10^{-2}$ and $\nu=10^{-8}$.  The epicenter block is indicated by the arrow.
}
\end{figure}

 We find that these two distinct recurrence times are actually originated from {\it two distinct types of seismic events\/}, which we call the type-I and the type-II events. The type-I event occurs with its epicenter lying just next to the rim of the rupture zone of the preceding event, and tends to be unilateral, {\it i,e,\/}, the rupture propagates only in one direction. By contrast, the epicenter of the type-II event lies in the interior of the rupture zone of the preceding event, and its rupture tends to propagate in both directions. The two types of events are illustrated in Fig.7 on the position versus the time plot.

 We confirm that the two peaks of the recurrence distribution of Fig.6 are indeed associated with these two types of events, {\it i.e.\/}, the peak at $T_1$ with the type-I event and the peak at $T_2$ with the type-II event. In our simulations, the type-I events are defined as events with its epicenter lying one block next to the rim of the rupture zone of preceding events, while all other mainshocks are regarded as type-II events. In Fig.6, we also show the ``dissolved'' recurrence time distributions for the type-I events and for the type-II events as defined above separately, which validates the above identification. 

\begin{figure}[ht]
\begin{center}
\includegraphics[scale=0.85]{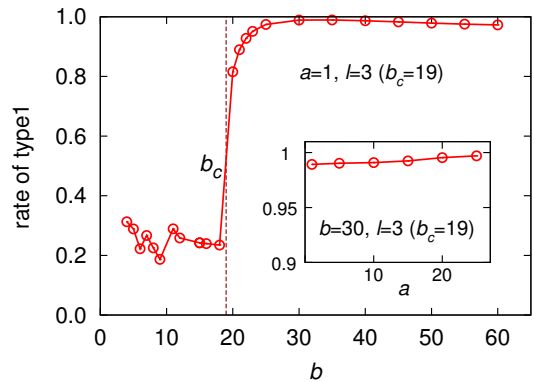}
\end{center}
\caption{
The $b$-dependence of the rate of the type-I event. The other parameter values are  $a=1$, $l=3$, $c=1000$, $v^*=10^{-2}$ and $\nu=10^{-8}$. The borderline value of $b$ separating the strong/weak frictional instability regimes, $b_c=19$, is indicated by the vertical dotted line. The inset shows the $a$-dependence of the rate of the type-I event at $b=b_c=19$.
}
\end{figure}

 We note that, as can be seen from Fig.7, the type-II event possesses a feature of the so-called ``asperity'' in that a nearly common area tends to rupture repeatedly with a nearly common block as an epicenter. The 1D BK model certainly contains a mechanism of stabilizing such an asperity-like event.

 A hint of such a stabilization mechanism might be obtained from the stress distribution just before and after the type-II event, an example of which is demonstrated in Fig.8. As can be seen from the figure, the higher stress before the mainshock is released after the mainshock in its rupture zone, while the stress drop near the epicenter remains modest. This is because, at the initial stage of the mainshock, the rupture has not been fully developed. In contrast, the stress drop  on the both sides of the epicenter region is significant, providing a relatively low-stress region surrounding the epicenter region of relatively high stress. Since the stress loading after the mainshock is uniform, the epicenter region of relatively high stress tends to be an epicenter of the next event, {\it provided\/} that it is not involved in other events which occur with its epicenter at some distant site outside of the rupture zone of the target event. Then, the low-stress region surronding the epicenter site serves to provide a stress ``trench'', preventing the epicenter site from being involved in the events propagated from the outer region. This mechanism works effectively especially in the present 1D model, stabilizing the type-II asperity-like event. By contrast, the mechanism is expected to be less effective in 2D simply due to the geometrical reason: the possible paths of the rupture propagation from outside could be far richer in 2D than in 1D so that the high stress state at the epicenter region tends to be more vulnerable to the rupture invasion from outside. If so, the type-II event would be more eminent in 1D than in 2D.

 Note that the asperity-like character of the type-II event is self-generated from the completely homogenous evolution law and homogenous material parameters. In seismology, the asperity-like events are usually attributed to the spatial inhomogeneity of the earthquake fault, {\it i.e.\/}, the asperity is considered to a special spot with a special geography or special material parameters distinct from other places. Our present result demonstrates, on the other hand, that {\it the completely homogenous system, at least in material parameters describing its equation of motion and constitutive law, still can self-generate asperity-like phenomena via its dynamical evolution\/}. Example of similar self-generated asperity-like phenomena in a spatially homogeneous setting was also reported in certain 2D coupled map lattice model \cite{Kotani, Kawamura2010}. Of course, the asperity-like type-II event sequence in the homogenous model does not last permanently. It is  interrupted at a certain stage, but could continue over many events, say, ten times. 

 In  Fig.9, the rate of the type-I events among all events is plotted versus the parameter $b$. In the strong frictional instability regime of larger $b$, most of the events are type I so that the $T_2$-peak of Fig.6 originated from the type-II events is faint. Especially for $b\gtrsim 30$, almost all events are type I. In the weak frictional instability regime of smaller $b$, on the other hand, the type-II events are dominant so that the $T_1$-peak is hardly discernible. Near the border $b\simeq b_c$, the type-I/type-II ratio exhibits a pronounced increase as $b$ is increased across $b_c$. Such a dominance of either type-I/II event for $b>b_c$ or $<b_c$ might explain the changeover observed in several observables. For example, the change in the form of the magnitude distribution shown in Fig.1 is attributed to the observation that the type-II event possesses a pronounced characteristic property with an eminent single-peaked distribution, while the type-I event tends to be less characteristic with a flat distribution spanning from smaller to larger events. As can be seen from the inset of Fig.9, the type-I/II ratio hardly depends on $a$. If one combines Fig.9 with Fig.4(c), one sees that the rupture length $L_0$ tends to be longer in the type-II events than in the type-I events by, say, a factor of two. This may simply reflect the fact that the type-I event is unilateral while the type-II one is bilateral. 

\begin{figure}[ht]
\begin{center}
\includegraphics[scale=0.85]{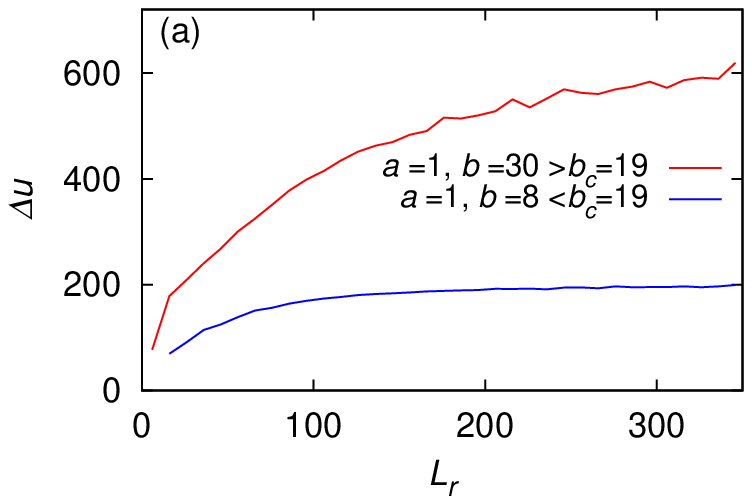}
\includegraphics[scale=0.85]{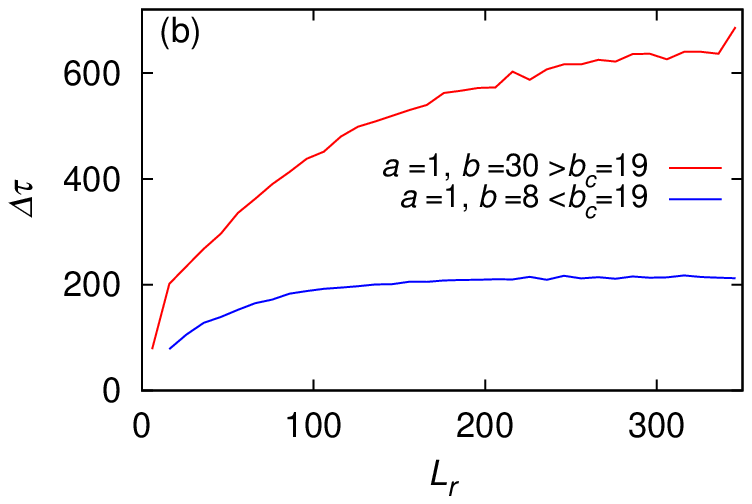}
\end{center}
\caption{
The rupture-length $L_r$ dependence of (a) the mean slip amount $\Delta u$, and of (b) the mean stress drop $\Delta \tau$. The $b$-value is either $b=30>b_c=19$ in the strong frictional instability regime (red), and  $b=8<b_c$ in the weak frictional instability regime (blue). The other parameter values are  $a=1$, $l=3$, $c=1000$, $v^*=10^{-2}$ and $\nu=10^{-8}$.
}
\end{figure}

 In Fig.10(a), the mean slip amount of blocks involved in an event is shown as a function of the rupture-zone size $L_r$ for the case of $b=30\geq b_c=19$ and of $b=8\leq b_c$, each corresponding to the strong and the weak frictional instability regimes. A similar plot is given also for the mean stress drop in Fig.10(b). The mean stress drop is defined here as the difference between the elastic forces at the onset and at the end of a given event, averaged over all blocks involved in this event, where the (dimensionless) elastic force at a given block $i$ is given by $\nu t-u_i+l^2(u_{i+1}-2u_i+u_{i-1})$. As can be seen from these figures, both the mean slip and the mean stress drop increase monotonically with $L_r$, and eventually tend to saturate taking nearly $L_r$-independent values. The tendency is more eminent in the weak frictional instability regime of $b<b_c$.

\begin{figure}[ht]
\begin{center}
\includegraphics[scale=0.85]{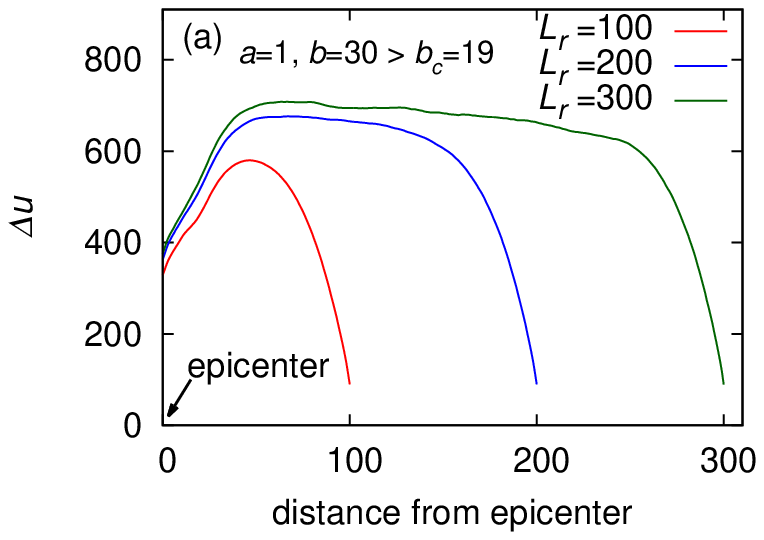}
\includegraphics[scale=0.85]{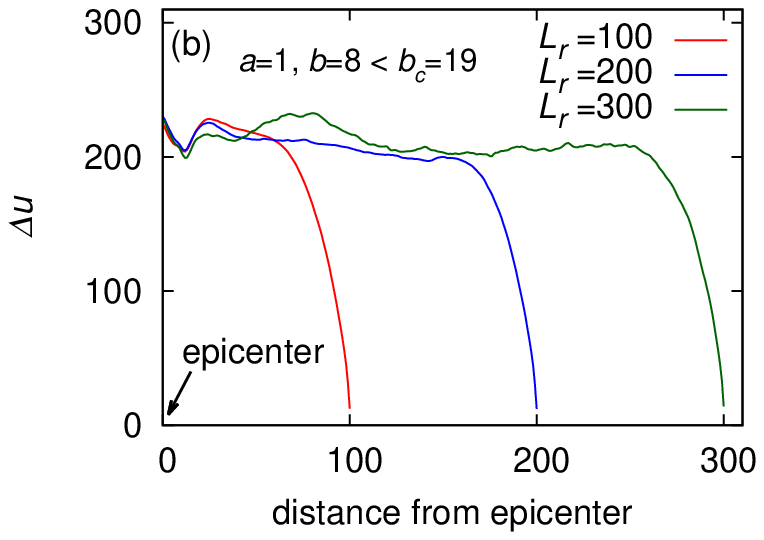}
\end{center}
\caption{
The mean slip amount during the rupture propagation of a mainshock is plotted versus the block position measured from the epicenter block.  The $b$-value is either (a) $b=30>b_c=19$ in the strong frictional instability regime, or (b) $b=8<b_c$ in the weak frictional instability regime. The other parameter values are  $a=1$, $l=3$, $c=1000$, $v^*=10^{-2}$ and $\nu=10^{-8}$. Events which propagate with exactly the distance $L_r=100, 200$ and 300 in their longer direction are collected, and the data are averaged over the events satisfying this condition.
}
\end{figure}
\begin{figure}[ht]
\begin{center}
\includegraphics[scale=0.85]{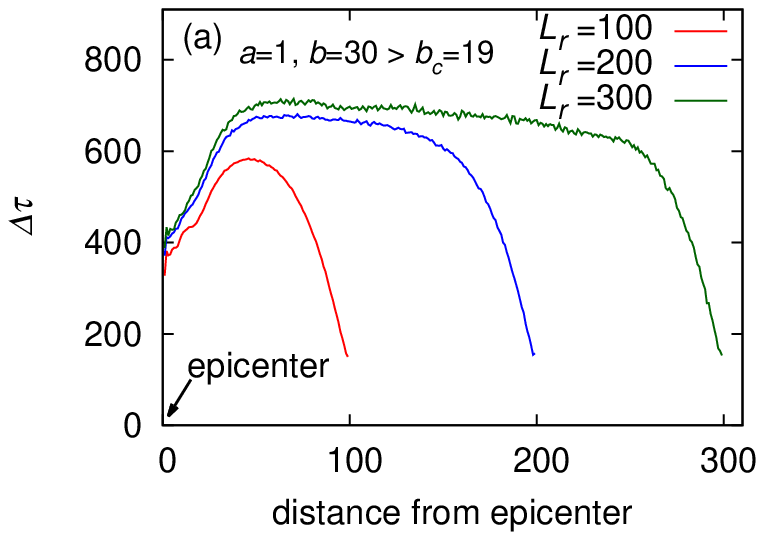}
\includegraphics[scale=0.85]{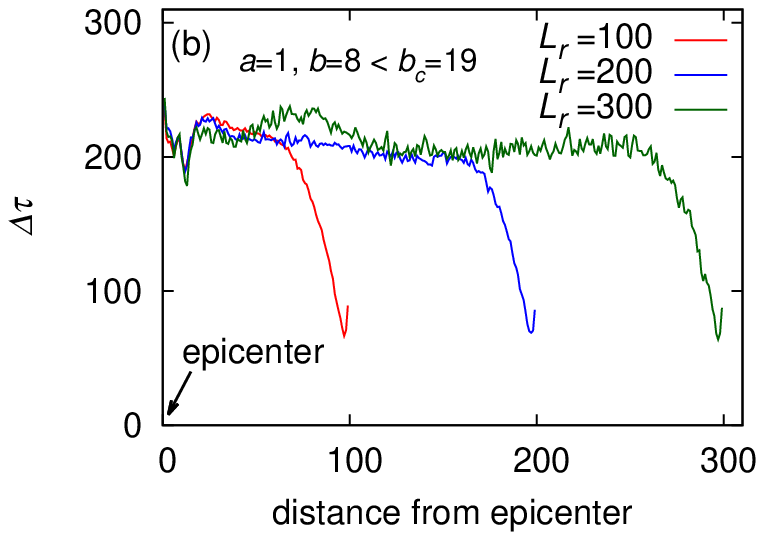}
\end{center}
\caption{
The mean stress drop during the rupture propagation of a mainshock is plotted versus the block position measured from the epicenter block.  The $b$-value is either (a) $b=30>b_c=19$ in the strong frictional instability regime, or (b) $b=8<b_c$ in the weak frictional instability regime. The other parameter values are  $a=1$, $l=3$, $c=1000$, $v^*=10^{-2}$ and $\nu=10^{-8}$. Events which propagate with exactly the distance $L_r =100, 200$ and 300 in their longer direction are collected, and the data are averaged over the events satisfying this condition.
}
\end{figure}

 It has been known for some time that in real seismicity the stress drop tends to be nearly independent of the event size taking a value around 3 MPa, though with rather large dispersions between 0.03 $\sim$ 30 MPa \cite{Scholz2002}.

 By contrast, if the stress drop is to be size-independent, the standard elastic theory would expect the mean slip being proportional to the rupture length $L_r$. As we shall show in \S IV  below, the event-size independence of the mean slip observed here for the discrete BK model actually persists in its continuum limit. Hence, the saturation of the mean slip with respect to the event size $L_r$ is not just due to the discreteness of the BK model, but is an essential property of the model construction. Presumably, this would be related to the way of the plate loading in the BK model, where the blocks constituting a deformable fault layer are directly pulled by the contingent moving rigid plate via the elastic springs, whereas, in the standard elastic continuum model, the plate loading is applied infinitely apart from the fault layer. Such setting implicitly assumed in the 1D BK model is expected to apply to long mature faults with its length much longer than the seismoginic-zone width. Such setting is also the one assumed in the so-called ``$W$-model'' \cite{Scholz1982,Ramanowicz}, which also predicts the saturation of the slip amount for very long ruptures. The scaling relation between the mean slip amount and the rupture length for natural faults has long been discussed \cite{Scholz1982,Ramanowicz,Mai,ShawScholz,Scholz2002,Manighetti,Blaser}, where some reported that the mean slip amount tended to be size independent for very large events with $L_r$ much longer than the seismogenic-zone width $W$ \cite{Mai,ShawScholz,Scholz2002,Manighetti}.

\begin{figure}[ht]
\begin{center}
\includegraphics[scale=0.85]{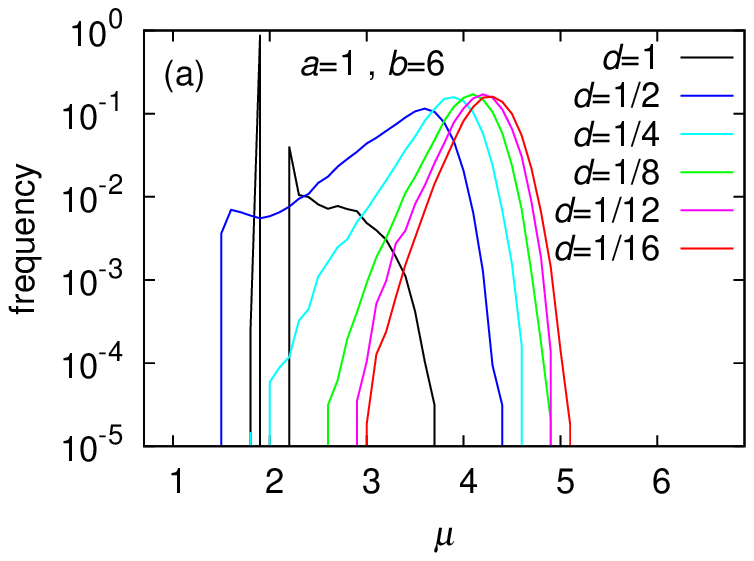}
\includegraphics[scale=0.85]{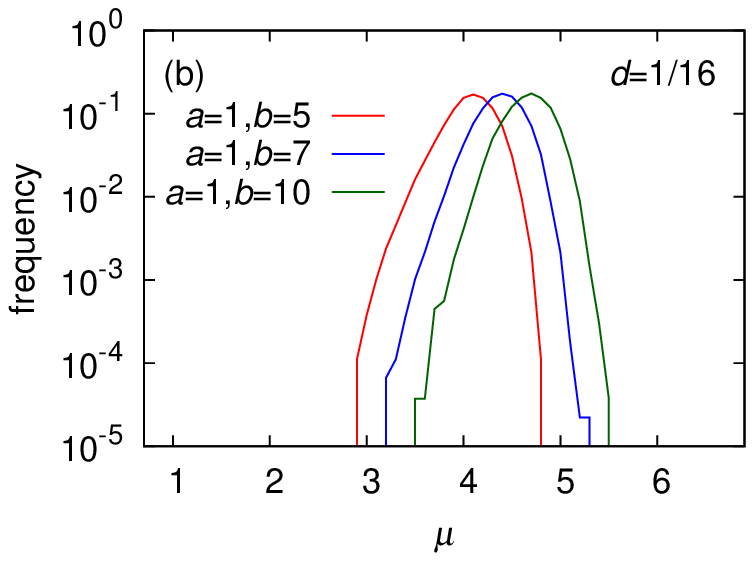}
\end{center}
\caption{
(a) The block-size $d$ dependence of the  magnitude distribution. The parameters are $a=1$, $b=6$, $c=1000$, $v^*=10^{-2}$ and $\nu=10^{-8}$. The $d\rightarrow 0$ limit corresponds to the continuum limit where $l=1/d$. The system size is $N/d$ with $N=800$. (b) The magnitude distribution near the continuum limit ($d=1/16$) for various $b$-values. The other parameters are the same as in (a).
}
\end{figure}

 Additional information can be obtained by looking at the manner how the rupture propagates during the mainshock. Thus, we show in Fig.11 the mean slip during the rupture propagation of a mainshock plotted versus the block position measured from the epicenter block. The $b$-value is either $b=30$ in the strong frictional instability regime (a), or $b=8$ in the weak frictional instability regime (b). Events which propagate exactly of the distance $L_r =100, 200$ and 300 in the longer direction are collected, and the data are averaged over the events satisfying this condition. Similar plots are also given for the mean stress drop in Fig.12 for the same parameters sets as Fig.11.

 As can be seen from Figs.11 and 12, both the mean slip and the mean stress drop tend to reach a constant plateau value except at the beginning and at the end of the rupture. Such a plateau-like behavior is more pronounced in the weak frictional instability regime, where the plateau-like behavior immediately sets in event near the epicenter block, presumably due to the accompanying  nucleation process in this regime \cite{Ueda2014,Ueda2015}. The important observation here is that this plateau value is independent of the event size $L_r$, except for smaller events not exhibiting a plateau behavior. Such a plateau-like behavior independent of the event size $L_r$ immediately explains the reason why the mean stress drop and the mean slip amount becomes nearly independent of the rupture length except for smaller events, especially in the weak frictional instability case. We note in passing that the plateau-like behavior might be originated from the pulse-like propagation of the rupture front in a mainshock, which tends to flatten the stress state within the rupture zone.

\begin{figure}[ht]
\begin{center}
\includegraphics[scale=0.85]{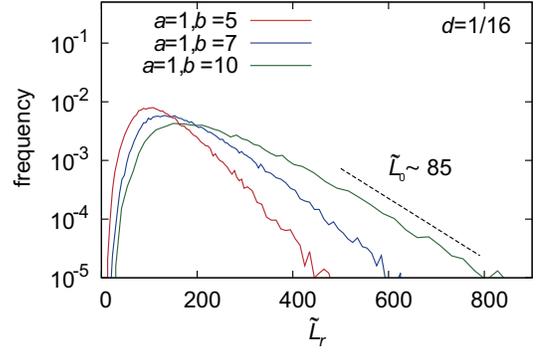}
\end{center}
\caption{
The rupture-length $\tilde L_r=L_rd=L_r/l$ distribution near the continuum limit ($d=1/16$) for various $b$-values. The other parameter values are $a=1$, $c=1000$, $v^*=10^{-2}$ and $\nu=10^{-8}$.
}
\end{figure}
\begin{figure}[ht]
\begin{center}
\includegraphics[scale=0.85]{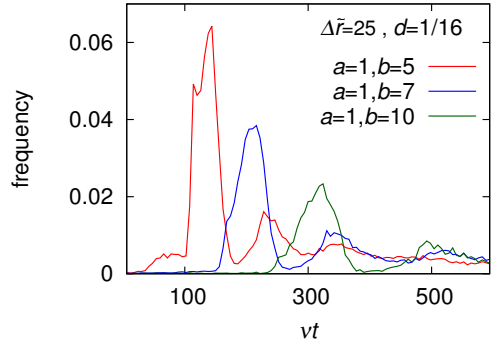}
\end{center}
\caption{
The mainshock recurrence-time distribution near the continuum limit ($d=1/16$) for various $b$-values. The other parameter values are $a=1$, $c=1000$, $v^*=10^{-2}$ and $\nu=10^{-8}$.
}
\end{figure}

\section{The continuum limit}

 In this section, we examine the continuum limit of the discrete BK model. The continuum limit corresponds to making the block size to be infinitesimally small, $d\rightarrow 0$, simultaneously making the system infinitely rigid $l\rightarrow \infty$ so that $d=1/l$ \cite{MoriKawamura2008c}. The dimensionless distance $x$ between the block $i$ and $i^{\prime}$ is given by
\begin{equation}
x=|i-i^{\prime}|d=\frac{|i-i^{\prime}|}{l}.
\label{continuum}
\end{equation}
As discussed in Ref.\cite{MoriKawamura2008c}, the equation of motion in the continuum limit is given  in the dimensionful form by
\begin{equation}
\frac{{\rm d}^2U}{{\rm d}t'^2} = \omega^2(\nu't'-U) + \xi^2 \frac{{\rm d}^2U}{{\rm d}x^2} - \Phi' ,
\label{continuumeq}
\end{equation}
where $U(x,t')$ is the displacement at the position $x$ and the time $t'$, $\Phi'$ is the friction force per unit mass, while $\omega$ and $\xi(\simeq v_s)$ are the characteristic frequency and  the characteristic wave-velocity, respectively.

\begin{figure}[ht]
\begin{center}
\includegraphics[scale=0.85]{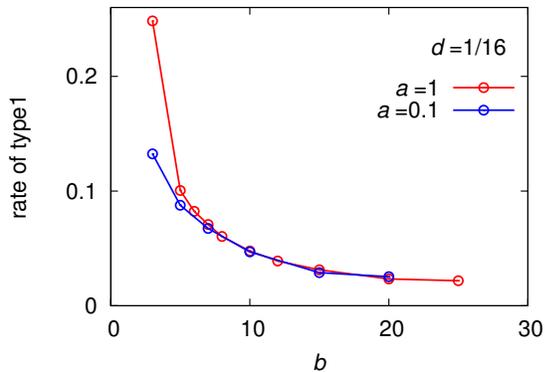}
\end{center}
\caption{
The $b$-dependence of the rate of the type-I event near the continuum limit ($d=1/16$) for $a=1$ and 0.1. The other parameter values are $c=1000$, $v^*=10^{-2}$ and $\nu=10^{-8}$.
}
\end{figure}

 The length unit scaling the block size is then $\xi/\omega$, while the length unit scaling the block displacement is the characteristic slip distance ${\mathcal L}$. 

 The continuum limit of the BK model necessarily lies in the weak frictional regime since $b_c=2l^2+1 \rightarrow \infty$ in this limit. Hence, the statistical properties of the model in its continuum limit should generally be those of the weak frictional instability regime with enhanced characteristic properties. 

  In order to examine the convergence toward the continuum limit $d\rightarrow 0$, we show in Fig.13(a) the magnitude distribution with systematically varying the block size $d$ from $d=1$ corresponding to the original BK model to smaller values down to $d=1/16$. The parameters are taken $a=1$, $b=6$, $c=1000$. As can be seen from the figure, the computed magnitude distribution tends to converge  as $d$ is taken smaller, approaching a limiting form. In fact, the convergence appears reasonably good already at $d=1/16$. Note that, although the initial ($d=1$) choice of the parameter lies in the {\it strong\/} frictional instability regime, the one emerging in the continuum limit resembles that of the weak frictional instability with an enhanced characteristic feature.

 In Fig.13(b), we show the magnitude distribution functions computed at $d=1/16$, expected to be close to the continuum limit, with varying the $b$-value as $b=5,7$ and 10. As expected from the fact that the continuum limit always lies in the weak frictional instability regime irrespective of the $b$-value, the computed magnitude distributions are always single-peaked with an enhanced characteristic feature irrespective of their $b$-value.

 In Fig.14, we show on a semi-log plot the rupture-length distribution computed at $d=1/16$ expected to be close to the continuum limit, with varying the $b$-value. As in the original model with $d=1$, the tail of the distribution exhibits a near-linear behavior corresponding to the exponential behavior, yielding a characteristic length scale associated with the inverse slope $\tilde L_0=L_0d=L_0/l$ in the continuum limit. Note that the length $\tilde L_0$ is measured here as has been given in eq.(\ref{continuum}) above. In the dimensionless unit, $\tilde L_0$ is around several tens, increasing with $b$. The result indicates that, in the continuum limit of the BK model, there exists a characteristic length scale for the mainshock rupture length. Recalling the length unit here to be $v_s/\omega \sim 2$ [km], this characteristic length scale may roughly be estimated to be $\tilde L_0 \sim 100$ [km]. If one literally translates the result into the real world, it means that, an event in a hypothetical infinite uniform fault obeying the RSF law, the mainshock rupture length cannot be indefinitely large, with a characteristic length scale of, say, $\sim $100 [km]. As a consequence of this exponential behavior of the rupture length, the occurrence probability of the events of, say, $\tilde L_r\gtrsim 10\tilde L_0$ is quite low, $\sim 0.005\%$, suggesting the practical upper limit of the rupture length of earthquakes being, say, $10\tilde L_0 \sim 1000$ [km]. Interestingly, this upper limit comes close to the rupture length of 1960 Chile Earthquake.

 In Fig.15, we show the mainshock recurrence-time distribution computed at $d=1/16$ close to the continuum limit, with varying the $b$-value. The pronounced peak structure corresponding to the occurrence of the next (second-next, ... {\it etc\/}.) mainshock persists as observed in the original model with $d=1$ in the weak frictional regime. It indicates the near periodic recurrence of mainshocks. The second and further peaks arise because the next event sometimes happens to be missed due to the somewhat arbitrary condition of the ``vicinity'' $\Delta r$, and the second-next event is counted as the next event. The computed recurrence-time distribution resembles the one estimated for large events at natural faults, in that the distribution exhibits a single peak at a characteristic magnitude \cite{Nishenko, Sykes}.

\begin{figure}[ht]
\begin{center}
\includegraphics[scale=0.85]{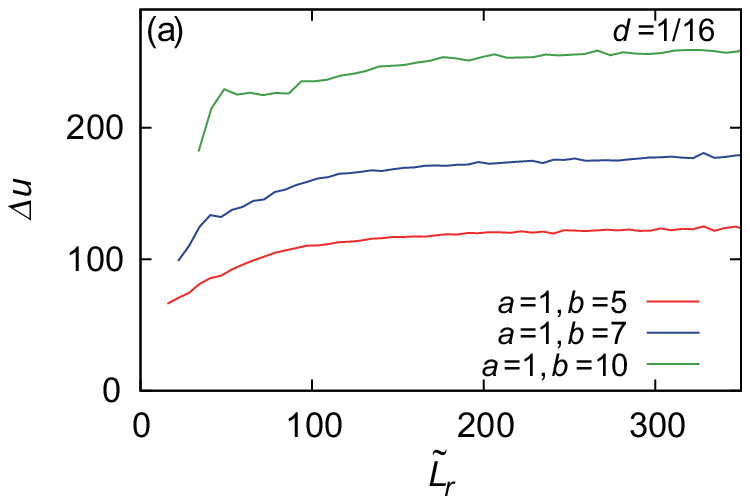}
\includegraphics[scale=0.85]{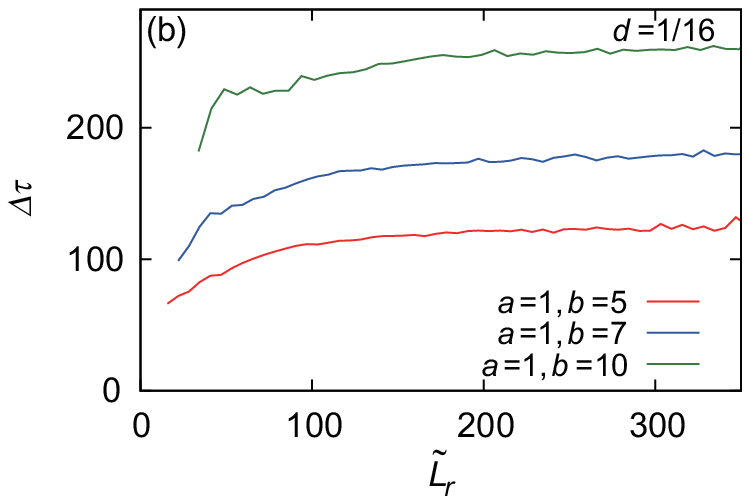}
\end{center}
\caption{
The rupture-length dependence of (a) the mean slip amount, and  (b) the mean stress drop,  which represents the continuum limit of Fig.10. The $b$-value is $b=5,7,10$, while the other parameters are $a=1$, $c=1000$, $v^*=10^{-2}$ and $\nu=10^{-8}$.
}
\end{figure}

  In Fig.16, we show the $b$-dependence of the rate of the type-I event computed at $d=1/16$ close to the continuum limit for the cases of $a=0.1$ and 1. Since the region lies in the weak frictional instability regime of the original model with $d=1$, most of the events should be of type-II so that the computed ratio stays rather small, being less than 0.1 for most of the parameter values. Yet, the type-I ratio tends to increase as $b$ gets smaller. There could be two factors in this increase. First, for $b<1$, the model exhibits a stationary creep-like behavior, no longer accompanying the standard high-speed rupture of a mainshock. Second, for $a\gtrsim b$, the mainshock changes its character, accompanied by a slow afterslip, and for $a$ significantly greater than $b$, the high-speed rupture no longer occurs but SSE occurs instead. If so, the change of behavior is expected as $b\ (>a)$ approaches $a$. Indeed,  in Fig.16, the rapid increase of the type-I rate occurs at a larger $b$-value for $a=1$ than for $a=0.1$. In other words, in the parameter region of $a\gtrsim b$ and/or $b<1$, the basic character of the events changes where the slow-slip behaviors tend to become dominant, significantly modifying the statistical properties. We leave this issue of the slow slip of the BK model in the future publication.

   In Fig.17, we show the rupture-length $\tilde L_r$ dependence of the mean slip amount (a), and of the mean stress drop (b) computed at $d=1/16$ close to the continuum limit, with varying the $b$-value as $b=5,7$ and 10. It corresponds to the continuum limit of Fig.10 of the original model with $d=1$. As expected, the obtained behavior is similar to the one observed in the original model in the weak frictional instability regime. In particular, both the mean stress drop and the mean slip amount tend to be independent of the rupture length except for smaller events.

\section{Summary and discussion}

Statistical properties of the one-dimensional spring-block (Burridge-Knopoff) model of earthquakes obeying the rate and state dependent friction law are studied by extensive numerical computer simulations. The quantities computed include the magnitude distribution, the rupture-length distribution, the mainshock recurrence-time distribution, the seismic time correlations before and after the mainshock, the mean slip amount, and the mean stress drop at the mainshock, {\it etc\/}.

 The statistical properties turned out to differ considerably depending on whether the system is either in the weak or the strong frictional instability regime, each corresponding to $b>b_c\ (=2l^2+1)$ or $b<b_c$, where $b$ is the frictional weakening parameter and $l$ is the elastic stiffness parameter. In the weak frictional instability regime, seismic events generally tend to possess enhanced characteristic features. The magnitude distribution, for example, changes its character depending on whether in the strong or the weak frictional instability regime: the distribution is eminently single-peaked in the weak frictional instability regime whereas tends to be flat in the strong frictional instability regime \cite{Ueda2014,Ueda2015}.

 Large events of the model can be classified into the two categories which we call type-I and II. The type-I event occurs with its epicenter located at the rim of the rupture zone of the previous event, and tend to be unilateral propagating mainly in only one direction. On the other hand, the type-II event resembles the asperity-like earthquake with its epicenter located in the interior of its rupture zone and recur near periodically, the rupture propagating into both directions. We observed that in the strong frictional instability regime large events are dominated by the type-I events, while in the weak frictional instability regime by the type-II events. The difference in the statistical properties in the strong/weak frictional instability regimes is understandable as the difference in the character of the type-I/II events. In particular, an enhanced characteristic feature observed in the weak frictional instability regime is originated from the enhanced characteristic feature of the type-II events.

 One interesting finding of our simulation is that the distribution of the rupture length $L_r$ exhibits an exponential behavior at larger sizes, $\approx \exp[-L_r/L_0]$ with a characteristic ``seismic correlation length'' $L_0$, both in the strong and the weak frictional instability regimes, indicating the existence of an intrinsic length scale associated with the mainshock size. $L_0$ is around $10\sim100$ lattice spacings, though somewhat $b$-dependent.

 We also studied the seismic time correlation before and after the time events, to examine whether the model exhibits a foreshock/aftershock sequence. Except for the gradual increase of the seismic activity toward the next mainshock at the time scale of the mainshock interseismic period, no clear signature of foreshock/aftershock sequences is observed in the model. In particular, the model does not exhibit a foreshock/aftershock sequence obeying the Omori (inverse Omori) law. This absence might partly be due to the one-dimensional feature of the model. It might be interesting to investigate the corresponding time correlations in higher dimensions.

 The continuum limit of the model is then examined, by systematically taking a finer block size. As discussed in section II, the continuum limit of the BK models differs from the standard elastodynamic model in that the characteristic time scale $\omega^{-1}$ has been introduced in its equation of motion. In the continuum limit, the model is expected to lie in the weak frictional instability regime irrespective of its parameter values. Indeed, we confirmed that this expectation was fully met. The event in the continuum limit of the 1D BK model exhibits pronounced characteristic features, corroborating the argument by Rice \cite{Rice}.

Meanwhile, in the parameter range of $b\lesssim 1$ or $a\gtrsim b$, the slow-slip phenomena come into play, considerably changing the character of seismic events. We will deal with such slow-slip regime in a separate paper.

 The properties of the BK model under the RSF law are sometimes considerably different from those of the well-studied BK model under the pure velocity-weakening law employed in most of the previous simulations on the model. Namely, characteristic features tend to be more enhanced in the RSF-law model than in the pure velocity-weakening-law model. This is presumably due to the fact that the RSF law possesses an intrinsic length scale in it, the characteristic slip distance, whereas the pure velocity-weakening law does not possess such a length scale.

 Finally, we wish to discuss possible implications of our present results to real seismicity, by providing rough estimates of various characteristic numbers, on the basis of the estimates of typical time and length scales given in Ref.\cite{Ueda2014}.  Of course, these should be taken only as rough estimates since our model itself is a very crude one. The characteristic rupture length $L_0$ of seismic events has been estimated to be $\sim100$ [km]. This means that event at a hypothetical homogenous infinite fault, events cannot be indefinitely large, say, $L_r\lesssim 10L_0$, the upper limit being of order several hundreds till a thousand kilometers. Interestingly, this upperlimit comes close to the rupture length of 1960 Chile Earthquake of $\sim 1000$ [km]. Events in the continuum limit tend to possess a character of asperity event (type-II in our notation) even in the completely homogeneous parameter setting, and tend to repeat quasi-periodically. The typical recurrence time can be estimated from Fig.15 to be 100$\sim$300 in dimensionless units, which, in the dimensionful number, corresponds to a few hundred years with the typical $\nu$-value of order a few [cm] per year. The typical slip amount might be estimated from Fig.17(a) to be 100$\sim$300 in dimensionless units, which, in the dimensionful number, corresponds to a few [m]. These numbers seem quite reasonable ones expected for large interplate earthquakes occurring at a mature interplate fault.

 Another interesting observation of the present study is that not only the mean stress drop but also the mean slip tends to be rupture-length ($L_r$) independent for larger events. Such a saturation of $L_r$ is similar to the one expected in the so-called ``$W$-model'', which was supported by large strike-slip earthquakes at natural faults  \cite{Mai,ShawScholz,Scholz2002,Manighetti}. Indeed, our Fig.17 suggests that such a saturation occurs around $\tilde L_r\gtrsim 100$ in the dimensionless units, which corresponds in real seismicity to large events of $L_r\gtrsim 200$ [km]. Interestingly, such a saturation of the mean slip amount against the rupture length was indeed reported. For example, Fig.1(a) of Ref.\cite{Manighetti} suggests that the slip amount tends to saturate for longer rupture length of $L_r\gtrsim 100-200$ [km].

 In this way, an important message from the present study is that, {\it at least in a mature homogeneous fault obeying the RSF law, events tend to be eminently characteristic\/} with characteristic length, time and energy scales. Our present model is 1D rather than 2D, which might over-emphasize the characteristic feature of the associated seismicity. Yet, our preliminary calculation on the corresponding 2D BK model suggests that the model keeps enhanced characteristic features in the continuum limit even in 2D. This observation would mean that, at least when one looks at a single mature homogenous fault, say, Nankai trough, earthquakes might be strongly characteristic, with peaked distributions in various observables. Indeed, such model observations appear to be supported by seismic observations on mature interplate faults \cite{Davison, Wesnousky, Sykes, Ishibe}.

 Of course, if ones looks at real seismicity, things often could be much more complex and erratic. Earthquake statistics is often characterized by power-laws without any characteristic scales as seen in, {\it e.g.\/}, the celebrated GR law. Then, an emerging,  highly important fundamental question would be what is the true origin of the observed ``complexity'' and apparently ``critical'' (power-law) behavior of real earthquakes. The answer to this question might not necessarily be unique. 

 One possible factor might be the ``inhomogeneity''. This has already be seen even in the present homogenous model in the form of the block discreteness. We have observed that the things tend to be more characteristic as one approaches the continuum limit. Enhanced discreteness drives the system toward the strong frictional instability regime where the things tend to be more erratic or critical, as emphasized by Rice many years ago \cite{Rice}. Of course, a real fault cannot be completely homogeneous even at a single fault, and there could be various forms and levels of inhomogeneity.

 Concerning the GR law, one plausible scenario of its origin might be the following: if one looks at events occurring at a single mature interplate fault, the magnitude distribution might indeed deviate from the GR law, possibly with a peak or some structure appearing at a magnitude value characteristic of that fault, whereas, if one takes an average over many different faults with different material parameters as is usually done in taking statistics, the characteristic magnitude scales compensate with each other, eventually leading to the unpeaked distribution apparently without any characteristic magnitude scale \cite{Davison, Wesnousky, Turcotte, Kawamura2012}. This is very different from the original SOC mechanism of producing the seismic ``criticality'', but a similar mechanism of producing power-laws or scale-invariance has been known in statistical physics and solid-state physics, especially in random and inhomogeneous systems, {\it e.g.\/}, the glass (spin glass, structural glass, {\it etc.}) problem \cite{Anderson}. 

 As the constitutive law, we have used the RSF law, now standard in seismology. Our knowledge of the friction law, however, is still limited, especially in the high-speed regime, so that there always remains a possibility that the deviation from the RSF law gives the resulting earthquake events certain critical features. Indeed, our previous simulations employing the velocity-weakening friction law yielded the seismic events with more enhanced critical features \cite{MoriKawamura2005, MoriKawamura2006, Kawamura2012}. 

 In any case, we hope that the present results on the 1D BK model under the RSF law would provide a useful reference in understanding the complex earthquake phenomena. Further extensions would be desirable in several directions. One is an extension to 2D. The 2D model possesses a richer geometry, and might modify some part of the present results. The second might be to take account of the degrees of freedom along the perpendicular-to-fault direction into the model. In the BK model, while the deformable fault layer is represented by an assembly of blocks, the perpendicular degrees of freedom are largely suppressed or simplified in that the block assembly is directly pulled by the moving rigid plate via only one layer of elastic springs. Although such a simplification might work for a mature interplate fault with a well-developed LVFZ, in order to evaluate the generality of the properties of the BK model, it might be interesting to examine the effect of the perpendicular-to-fault degrees of freedom neglected in the original BK model by appropriately extending it. The third might be to study the effect of inhomogeneity of various forms and levels, a part of which we discussed above. The last is to study the slow-slip phenomena within the BK model. Indeed, our preliminary simulations suggest that the SSE could be describable even within the BK model. Then, it would be highly intersting to study the inter-relation between the SSE and the usual high-speed rupture within a single framework of the simple BK model to get deeper insight into general seismicity.

 As such, the BK model, in spite of its long history and its apparent simplicity, still remains to be a fruitful model involving rich physics to be uncovered. It provides us a useful reference point in understanding basic physical processes underlying apparently complex earthquake phenomena.

\begin{acknowledgments}
The authors are thankful to H. Kanamori, N. Kato, N. Hatano, T. Uchide, and T. Okubo for useful discussion. They are thankful to ISSP, the University of Tokyo for providing us with CPU time. This study is supported by a Grant-in-Aid for Scientific Research No.16K13851, and by the Ministry of Education, Culture, Sports, Science and Technology (MEXT) of Japan, under its Earthquake and Volcano Hazards Observation and Research Program.
\end{acknowledgments}

\end{document}